\documentclass
[superscriptaddress,secnumarabic,amssymb,amsmath,nobibnotes,aps,prd,showkeys,showpacs,nofootinbib,onecolumn,notitlepage,12pt]{revtex4-2}%
\usepackage{setspace}
\usepackage{color}
\usepackage{amsmath}
\usepackage{amsfonts}
\usepackage{verbatim}
\usepackage{amssymb}
\usepackage{graphicx,bm}
\usepackage{graphicx}
\usepackage{bbm}
\usepackage{amsmath}
\usepackage{amssymb}
\usepackage{bbm}
\usepackage{amssymb}
\usepackage{graphicx,bm}
\usepackage{graphicx}
\usepackage{bbm}
\usepackage{epstopdf}
\usepackage[caption=false]{subfig}
\usepackage{makecell}%
\providecommand{\U}[1]{\protect\rule{.1in}{.1in}}
\newcommand{\be}{\begin{equation}}
\newcommand{\ee}{\end{equation}}

\newcommand{\mincir}{\raise
-3.truept\hbox{\rlap{\hbox{$\sim$}}\raise4.truept\hbox{$<$}\ }}
\newcommand{\magcir}{\raise
-3.truept\hbox{\rlap{\hbox{$\sim$}}\raise4.truept\hbox{$>$}\ }}

\ifx\pdfoutput\relax\let\pdfoutput=\undefined\fi
\newcount\msipdfoutput
\ifx\pdfoutput\undefined\else
\ifcase\pdfoutput\else
\ifx\paperwidth\undefined\else
\ifdim\paperheight=0pt\relax\else\pdfpageheight\paperheight\fi
\ifdim\paperwidth=0pt\relax\else\pdfpagewidth\paperwidth\fi
\fi\fi\fi
\begin{document}
\title{Self-similar Cosmological Solutions in Symmetric Teleparallel theory:
Friedmann--Lema\^{\i}tre--Robertson--Walker spacetimes}
\author{N. Dimakis}
\email{nsdimakis@scu.edu.cn ; nsdimakis@gmail.com}
\affiliation{Center for Theoretical Physics, College of Physics,
Sichuan University, Chengdu 610064, China}
\author{M. Roumeliotis}
\email{microum@phys.uoa.gr}
\affiliation{Nuclear and Particle Physics section, Physics Department, University of
Athens, 15771 Athens, Greece}
\author{A. Paliathanasis}
\email{anpaliat@phys.uoa.gr}
\affiliation{Institute of Systems Science, Durban University of Technology, Durban 4000,
South Africa}
\affiliation{Departamento de Matem\'{a}ticas, Universidad Cat\'{o}lica del Norte, Avda.
Angamos 0610, Casilla 1280 Antofagasta, Chile}
\affiliation{Mathematical Physics and Computational Statistics Research Laboratory,
Department of Environment, Ionian University, Zakinthos 29100, Greece}
\author{P.S. Apostolopoulos}
\email{papostol@ionio.gr}
\affiliation{Mathematical Physics and Computational Statistics Research Laboratory,
Department of Environment, Ionian University, Zakinthos 29100, Greece}
\author{T. Christodoulakis}
\email{tchris@phys.uoa.gr}
\affiliation{Nuclear and Particle Physics section, Physics Department, University of
Athens, 15771 Athens, Greece}

\begin{abstract}
The existence of self-similar solutions is discussed in symmetric teleparallel
$f\left(  Q\right)  $-theory for a Friedmann--Lema\^{\i}tre--Robertson--Walker
background geometry with zero and non-zero spatial curvature. For the four
distinct families of connections which describe the specific cosmology in
symmetric teleparallel gravity, the functional form of $f\left(  Q\right)  $
is reconstructed. Finally, to see if the analogy with General Relativity holds, we discuss the
relation of the self-similar solutions with the asymptotic behaviour of more
general $f\left(  Q\right)  $ functions.

\end{abstract}
\maketitle
\date{\today}

\section{Introduction}

\label{sec1}

With the term of self-similar solutions we refer to a family of solutions with
the characteristic to map to itself after an appropriate scale of the
dependent or independent variables. In gravitational physics, self-similar
solutions of the Einstein field equations are mainly related with the
similarity solutions provided by the existence of a proper homothetic vector
field for the physical space \cite{bok1,Taub}. With the homothetic symmetry we
refer to the generator of the infinitesimal transformation in the physical
space which preserves the angles between the lines but not the scale. Previous
studies on exact solutions on homogeneous and inhomogeneous spacetimes
indicate that self-similar solutions correspond to asymptotic limits of more
general solutions \cite{as1,as2,as3,as4,as5}. Recall that self-similar
spacetimes cannot describe asymptotically flat or asymptotically spatially
compact geometries \cite{as6}.

Because of the importance of the self-similar solutions there is a plethora of
studies in the literature where the existence of a proper homothetic vector
field is investigated in various geometries. The conformal symmetries for the
Friedmann--Lema\^{\i}tre--Robertson--Walker (FLRW) spacetimes are investigated
in \cite{hv1}, where it is found that for a power-law scale factor a proper
homothetic vector field exists. Indeed, the exact solutions of ideal gas for a
spatially flat FLRW geometry or the Milne Universe, admit a proper homothetic
symmetry. The existence of homothetic vector field for a Bianchi I geometry
was studied in \cite{hv2a}. It was found that the Kasner solution as also the
Kasner-like solutions are self-similar solutions. An analysis of the
homothetic vector field in Bianchi III and Bianchi V can be found in
\cite{hv3}, while the four-dimensional stationary axisymmetric vacuum
spacetimes with a homothetic vector field were in detail investigated in
\cite{hv4}. On the other hand anisotropic and homogeneous self-similar exact
solutions for Bianchi VIII, Bianchi IX, and Bianchi VI$_{0}$ geometries with
tilted perfect fluid were studied in \cite{hv5,hvBianchiclassA, hv6}, while
Bianchi Class B spacetimes with a homothetic vector field was the subject of
study in \cite{hvequilBianciVIh, hv7}. Some inhomogeneous self-similar
solutions are presented in \cite{hv8}.

The fundamental invariant of Einstein's General Relativity (GR) is the Ricci
scalar $R$ defined by the symmetric Levi-Civita. Nevertheless more general
connections from that of the Levi-Civita have been used in gravitational
physics. Indeed, in teleparallelism the torsion scalar $\mathrm{T}$ defined by
the curvatureless Weitzenb\"{o}ck connection \cite{Weitzenb23} is the
geometric object which has been used to define gravity, leading to the
Teleparallel equivalent of General Relativity (TEGR)
\cite{Hayashi79,Maluf:1994ji}. Furthermore, from the non-metricity components
of a general connection the scalar $Q$ can be defined, where in the case of a
torsion-free and flat connection we end up with the Symmetric Teleparallel Equivalent
of General Relativity (STEG) \cite{Nester:1998mp}. Scalars $R$, $\mathrm{T}$
and $Q$ form the so-called geometrical trinity of gravity \cite{tr1}. The
gravitational Lagrangians, which are linear in each element of the trinity, lead
to the same gravitational theory. The latter equivalence is violated when
nonlinear terms of the geometric scalars are introduced in the gravitational Lagrangian.

The modification of the Einstein-Hilbert Action Integral is the simplest
geometric mechanism for the introduction of new degrees of freedom in order to
explain the observational phenomena \cite{mod1,Heisgen}. There exist a family
of modified theories of gravity known as $f\left(  X\right)  $-theory, where
the gravitational Lagrangian is a function $f$ of the geometric scalar $X$.
Usually $X$ is one element of the geometric trinity, such that when $f$ is a
linear function, GR (or TEGR or STEG) is recovered (respectively); of course
more complicated configurations can be constructed as well involving combinations of scalars \cite{Harko}. The
$f\left(  R\right)  $-theory of gravity introduced in \cite{fr} is a
fourth-order theory of gravity which can be written in a dynamical equivalent
form of the Brans-Dicke scalar field with zero Brans-Dicke parameter. Thus,
$f\left(  R\right)  $-theory is equivalent with a minimally coupled scalar
field under a conformal transformation \cite{Sotiriou}. $f\left(  R\right)
$-theory has been used to describe various epochs of the cosmological history,
such as the inflation \cite{star} or the late time acceleration \cite{odin2}.
Similarly the $f\left(  \mathrm{T}\right)  $-teleparallel theory of gravity
introduced as a geometric dark energy model \cite{Ferraro}, while it has been
widely applied in various gravitational configurations \cite{Jarv}, for more
details and applications of $f\left(  \mathrm{T}\right)  $-theory we refer the
reader to the recent review \cite{rew1}. Only recently the $f\left(  Q\right)
$-symmetric teleparallel theory \cite{f6} has drawn the attention of the
scientific society
\cite{ww0,ww1,ww2,ww3,ww4,ww5,ww6,ww7,ww8,ww9,ww10,ww11,ww12,ww13,ww14,ww15,ww16,ww17,ww18,ww19}.

In $f\left(  Q\right)  $-theory we make use of a flat connection pertaining
the existence of affine coordinates in which all its components vanish, that
is, turning covariant derivatives into partial (coincident gauge).
Consequently, in this theory it is possible to separate the inertial effects
from gravity. Thus the coincident gauge is always achievable through an
appropriate coordinate transformation. Recently, in \cite{nfq} the effects of
the use of different connections for the definition of the scalar $Q$ in the
dynamics of FLRW cosmology in $f\left(  Q\right)  $-theory was the subject of
study. For the FLRW spacetimes there exist four distinct families of
connections, compatible with the isometries of the FLRW metric
\cite{Hohmann,Heis2}, three for the spatially flat case and one when the
spatial curvature is present; for these connections exact solutions were
derived. In a later study, the effects of the different connections were
investigated in the case of static spherical symmetric spacetimes \cite{nfq2}.

In this work we are interested in the existence of self-similar solutions in
$f\left(  Q\right)  $-theory and on the effects of the different connections
on the existence of the homothetic symmetry vector for the case of a FLRW
background geometry. Self-similar solutions\ were investigated before in
$f\left(  R\right)  $-theory \cite{sm1,sm2} and in $f\left(  T\right)
$-teleparallel theory of gravity \cite{sm3}.

The structure of the paper is as follows: In Section \ref{sec2}, we present the
basic geometric elements of the $f\left(  Q\right)  $-symmetric teleparallel
theory and the different connections for the FLRW spacetime with or without
spatial curvature. In Section \ref{sec3}, we consider the spatially flat case
and we reconstruct closed-form expressions for the $f\left(  Q\right)  $
function, where self-similar solutions exist. The analysis for a non-zero
spatial curvature is presented in Section \ref{sec4}. Finally, in Section
\ref{sec5}, we summarize our results and draw our conclusions.

\section{Symmetric Teleparallel theory}

\label{sec2}

The metric $g_{\mu\nu}$ and the connection $\Gamma_{\;\mu\nu}^{\kappa}$ are
the basic dynamical objects in metric-affine gravitational theories. We define
the following fundamental tensors, the curvature $R_{\;\lambda\mu\nu}^{\kappa
}$, the torsion $\mathrm{T}_{\mu\nu}^{\lambda}$ and the non-metricity
$Q_{\lambda\mu\nu}$,
\begin{align}
R_{\;\lambda\mu\nu}^{\kappa}  &  =\frac{\partial\Gamma_{\;\lambda\nu}^{\kappa
}}{\partial x^{\mu}}-\frac{\partial\Gamma_{\;\lambda\mu}^{\kappa}}{\partial
x^{\nu}}+\Gamma_{\;\lambda\nu}^{\sigma}\Gamma_{\;\mu\sigma}^{\kappa}%
-\Gamma_{\;\lambda\mu}^{\sigma}\Gamma_{\;\mu\sigma}^{\kappa}\\
\mathrm{T}_{\mu\nu}^{\lambda}  &  =\Gamma_{\;\mu\nu}^{\lambda}-\Gamma
_{\;\nu\mu}^{\lambda}\\
Q_{\lambda\mu\nu}  &  =\nabla_{\lambda}g_{\mu\nu}=\frac{\partial g_{\mu\nu}%
}{\partial x^{\lambda}}-\Gamma_{\;\lambda\mu}^{\sigma}g_{\sigma\nu}%
-\Gamma_{\;\lambda\nu}^{\sigma}g_{\mu\sigma}.
\end{align}
In the latter expressions, the symbol $\nabla_{\mu}$ means covariant
derivative with respect to the affine connection $\Gamma_{\;\mu\nu}^{\kappa}$.
For a symmetric connection, as the one we consider in this work, the torsion
tensor vanishes, $\mathrm{T}_{\mu\nu}^{\lambda}=0$. Moreover, in symmetric
teleparallelism\thinspace\ the curvature tensor is also zero, that is
$R_{\;\lambda\mu\nu}^{\kappa}=0$, while for the non-metricity part we have
$Q_{\lambda\mu\nu}\neq0$.

The fundamental non-metricity scalar $Q$ of the symmetric teleparallel theory
is defined as%
\begin{equation}
Q=Q_{\lambda\mu\nu}P^{\lambda\mu\nu}, \label{defQ}%
\end{equation}
where $P_{\;\mu\nu}^{\lambda}$ is the non-metricity conjugate tensor expressed
as
\begin{equation}
P_{\;\mu\nu}^{\lambda}=-\frac{1}{4}Q_{\;\mu\nu}^{\lambda}+\frac{1}{2}%
Q_{(\mu\phantom{\lambda}\nu)}^{\phantom{(\mu}\lambda\phantom{\nu)}}+\frac
{1}{4}\left(  Q^{\lambda}-\bar{Q}^{\lambda}\right)  g_{\mu\nu}-\frac{1}%
{4}\delta_{\;(\mu}^{\lambda}Q_{\nu)}, \label{defP}%
\end{equation}
in which the contractions $Q_{\lambda}=Q_{\lambda\mu}%
^{\phantom{\lambda\mu}\mu}$, $\bar{Q}_{\lambda}=Q^{\mu}_{\phantom{\mu}\lambda
\mu}$ are introduced.

\subsection{$f\left(  Q\right)  $-theory}

In $f\left(  Q\right)  $-theory, the gravitational Lagrangian density is
defined by a generally non-linear function $f\left(  Q\right)  $ namelly
\cite{Zhao}
\begin{equation}
S=\frac{1}{2}\int d^{4}x\sqrt{-g}f(Q)+\int d^{4}x\sqrt{-g}\mathcal{L}%
_{M}+\lambda_{\kappa}^{\;\lambda\mu\nu}R_{\;\lambda\mu\nu}^{\kappa}%
+\tau_{\lambda}^{\;\mu\nu}\mathrm{T}_{\;\mu\nu}^{\lambda}, \label{action}%
\end{equation}
where, $g=\mathrm{det}(g_{\mu\nu})$, $\mathcal{L}_{M}$ is the matter fields'
Lagrangian density, and $\lambda_{\kappa}^{\;\lambda\mu\nu}$, $\tau_{\lambda
}^{\;\mu\nu}$ are Lagrange multipliers, whose variation impose the conditions
$R_{\;\lambda\mu\nu}^{\kappa}=0$ and $\mathrm{T}_{\;\mu\nu}^{\lambda}=0$.

The gravitational field equations of the $f\left(  Q\right)  $-theory, for the metric $g_{\mu\nu}$, are
\begin{equation}
\frac{2}{\sqrt{-g}}\nabla_{\lambda}\left(  \sqrt{-g}f^{\prime}(Q) P_{\;\mu\nu}
^{\lambda}\right)  -\frac{1}{2}f(Q)g_{\mu\nu}+f^{\prime}(Q)\left(  P_{\mu
\rho\sigma}Q_{\nu}^{\;\rho\sigma}-2Q_{\rho\sigma\mu}P_{\phantom{\rho\sigma}\nu
}^{\rho\sigma}\right)  =T_{\mu\nu}, \label{feq1a}%
\end{equation}
where now a prime denotes total derivative with respect to the variable $Q$,
that is, $f^{\prime}(Q)=\frac{df}{dQ}$ and $T_{\mu\nu}=-\frac{2}{\sqrt{-g}%
}\frac{\partial\left(  \sqrt{-g}\mathcal{L}_{M}\right)  }{\partial g^{\mu\nu}%
}$ is the energy-momentum tensor, which describes the matter components of the
gravitational fluid.

An equivalent way to write the field equations (\ref{feq1a}) is with the use
of the Einstein-tensor $G_{\mu\nu}$ such that \cite{Zhao},
\begin{equation}
f^{\prime}(Q)G_{\mu\nu}+\frac{1}{2}g_{\mu\nu}\left(  f^{\prime}%
(Q)Q-f(Q)\right)  +2f^{\prime\prime}(Q)\left(  \nabla_{\lambda}Q\right)
P_{\;\mu\nu}^{\lambda}=T_{\mu\nu}, \label{feq1}%
\end{equation}
where $G_{\mu\nu}=\tilde{R}_{\mu\nu}-\frac{1}{2}g_{\mu\nu}\tilde{R},$ with
$\tilde{R}_{\mu\nu}$ and $\tilde{R}$ being the Riemannian Ricci tensor and
scalar respectively which are constructed from the Levi-Civita connection. A
direct comparison with General Relativity can be perceived as the effect of an
effective energy momentum tensor
\begin{equation}
\mathcal{T}_{\mu\nu}=-\frac{1}{f^{\prime}(Q)}\left[  \frac{1}{2}g_{\mu\nu
}\left(  f^{\prime}(Q)Q-f(Q)\right)  +2f^{\prime\prime}(Q)\left(
\nabla_{\lambda}Q\right)  P_{\;\mu\nu}^{\lambda}\right]  . \label{Teff}%
\end{equation}
With the help of (\ref{Teff}), the resulting field equations can be written in
the simple form%
\begin{equation}
G_{\mu\nu}=\mathcal{T}_{\mu\nu}+\frac{1}{f^{\prime}(Q)}T_{\mu\nu}.
\end{equation}

In addition the variation of the Action Integral (\ref{action}) with respect
to the connection gives the field equations%
\begin{equation}
\nabla_{\mu}\nabla_{\nu}\left(  \sqrt{-g}f^{\prime}(Q)
P_{\phantom{\mu\nu}\sigma}^{\mu\nu}\right)  =0. \label{feq2}%
\end{equation}

From the definition of the effective energy momentum tensor, $\mathcal{T}%
_{\mu\nu}$, we observe that, if $f(Q)\propto Q$ the limit of General Relativity
is recovered since $\mathcal{T}_{\mu\nu}=0$. Moreover, when $Q=$const., Eq.
\eqref{feq2} leads to solutions of General Relativity with a cosmological
constant $\Lambda$.

The basic objects in the theory are the metric $g_{\mu\nu}$ and the connection $\Gamma_{\;\mu\nu}^{\lambda}$, for which, equations \eqref{feq1a} and \eqref{feq2} have to be solved respectively. Note also that these equations incorporate the effect of the variation of the lagrange multipliers. Therefore, for  whatever connection is obtained through the field equations, there can always be found a coordinate transformation $x\mapsto \tilde{x}$, under whose effect, the transformed connection becomes zero:
\begin{equation}
  \bar{\Gamma}^{\lambda}_{\;\mu\nu}(\tilde{x}) = \frac{\partial \tilde{x}^\lambda}{\partial x^\rho} \frac{\partial x^\eta}{\partial \tilde{x}^\mu} \frac{\partial x^\sigma}{\partial \tilde{x}^\nu} \Gamma^\rho_{\;\eta\sigma}(x) - \frac{\partial x^\rho}{\partial \tilde{x}^\nu} \frac{\partial x^\sigma}{\partial \tilde{x}^\mu} \frac{\partial^2 \tilde{x}^\lambda}{\partial x^\rho \partial x^\sigma} =0 .
\end{equation}
This stems from the two basic properties of $\Gamma_{\;\mu\nu}^{\lambda}$, being both symmetric and flat, $R_{\;\lambda\mu\nu}^{\kappa}=0$ \cite{Eisenhart,KoivistoLIT}.

Since trivializing the connection is a matter of a coordinate transformation, a possible strategy is to enforce \textit{a priori}  $\Gamma_{\;\mu\nu}^{\lambda}=0$ into the field equations; this is referred in the literature as the adoption of the coincident gauge. However, special care is needed when also making some particular ansatz for the metric. Assuming a specific type of $g_{\mu\nu}$ already consists a partial gauge fixing. So, it may happen that $\Gamma_{\;\mu\nu}^{\lambda}=0$, together with the ansatz for $g_{\mu\nu}$, over-restrict the system. An example of this can be seen in the simple case of a spatially flat FLRW space-time. If you write the metric in spherical coordinates, then the condition $\Gamma_{\;\mu\nu}^{\lambda}=0$ in the equations is no longer a gauge choice, but rather a restriction \cite{Zhao}. The problem is resolved if instead you consider the FLRW metric in Cartesian coordinates, which are compatible with having $\Gamma_{\;\mu\nu}^{\lambda}=0$.

In this work we do not assume blindly the coincident gauge. We rather explore the different possibilities that exist and which are compatible with the system of equations for a given form of the metric.

%Due to the fact that the curvature of the connection vanishes, that is
%$R_{\;\lambda\mu\nu}^{\kappa}=0$, there always exists a coordinate system in
%which the connection becomes zero, $\Gamma_{\;\mu\nu}^{\lambda}=0$,
%\cite{Eisenhart}. This is referred as the coincident gauge.

For the matter energy-momentum tensor, the constraint $T_{\phantom{\mu}\nu
;\mu}^{\mu}=0$ is still valid, which is the conservation law of mass. With
\textquotedblleft$;$\textquotedblright\ we denote the covariant derivative
with respect to the Christoffel symbols. Thus, the $\mathcal{T}%
_{\phantom{\mu}\nu;\mu}^{\mu}=0$ relation resulting of the equation
\eqref{feq2} for the connection, can be considered as a conservation law
for the theory \cite{Hisencons}.

\subsection{FLRW background geometry}

The FLRW line element in spherical coordinates reads
\begin{equation}
ds^{2}=-N(t)^{2}dt^{2}+a(t)^{2}\left[  \frac{dr^{2}}{1-kr^{2}}+r^{2}\left(
d\theta^{2}+\sin^{2}\theta d\phi^{2}\right)  \right]  , \label{genlineel}%
\end{equation}
where $k$ is the spatial curvature, $k=0$ describes a spatially flat universe,
$k=1$ describes a closed universe and $k=-1$ denotes an open universe.
Moreover, function $a\left(  t\right)  $ is the scale factor describes the
radius of the universe and $N\left(  t\right)  $ is the lapse function.
Without loss of generality we assume that $N\left(  t\right)  =1$. \ The FLRW
spacetimes admit a six dimensional symmetry group with generators%
\begin{equation}
\zeta_{1}=\sin\phi\partial_{\theta}+\frac{\cos\phi}{\tan\theta}\partial_{\phi
},\quad\zeta_{2}=-\cos\phi\partial_{\theta}+\frac{\sin\phi}{\tan\theta
}\partial_{\phi},\quad\zeta_{3}=-\partial_{\phi} \label{Kil1}%
\end{equation}
and
\begin{equation}%
\begin{split}
\xi_{1}  &  =\sqrt{1-kr^{2}}\sin\theta\cos\phi\partial_{r}+\frac
{\sqrt{1-kr^{2}}}{r}\cos\theta\cos\phi\partial_{\theta}-\frac{\sqrt{1-kr^{2}}%
}{r}\frac{\sin\phi}{\sin\theta}\partial_{\phi}\\
\xi_{2}  &  =\sqrt{1-kr^{2}}\sin\theta\sin\phi\partial_{r}+\frac
{\sqrt{1-kr^{2}}}{r}\cos\theta\sin\phi\partial_{\theta}+\frac{\sqrt{1-kr^{2}}%
}{r}\frac{\cos\phi}{\sin\theta}\partial_{\phi}\\
\xi_{3}  &  =\sqrt{1-kr^{2}}\cos\theta\partial_{r}-\frac{\sqrt{1-kr^{2}}}%
{r}\sin\theta\partial_{\phi}.
\end{split}
\label{Kil2}%
\end{equation}

In two independent recent studies, Hohmann \cite{Hohmann} and D'Ambrosio et.
al. \cite{Heis2} derived the general form of all compatible connections for
the line element (\ref{genlineel}) for the symmetric teleparallel theory by
enforcing on a generic connection the six Killing symmetries and the
requirement $R_{\;\lambda\mu\nu}^{\kappa}=0$. In what follows we briefly
summarize the results.

For the spatially flat spacetime, i.e. $k=0$, there are three compatible
connections. The common non-zero components of all three are:
\begin{equation}%
\begin{split}
&  \Gamma_{\;\theta\theta}^{r}=-r,\quad\Gamma_{\;\phi\phi}^{r}=-r\sin
^{2}\theta,\\
&  \Gamma_{\;r\theta}^{\theta}=\Gamma_{\;\theta r}^{\theta}=\Gamma_{\;r\phi
}^{\phi}=\Gamma_{\;\phi r}^{\phi}=\frac{1}{r},\quad\Gamma_{\;\phi\phi}%
^{\theta}=-\sin\theta\cos\theta,\Gamma_{\;\theta\phi}^{\phi}=\Gamma
_{\;\phi\theta}^{\phi}=\cot\theta.
\end{split}
\label{common}%
\end{equation}
However, they do differ in the way a free function of time enters in some of
their other components. The first connection, named hereafter $\Gamma_{1}$,
has only one additional non-zero component
\begin{equation}
\Gamma_{\;tt}^{t}=\gamma(t), \label{con1}%
\end{equation}
where $\gamma(t)$ is a function of the time variable $t$.

The second connection, $\Gamma_{2}$, possesses the - additional to
\eqref{common} - non-zero components
\begin{equation}
\Gamma_{\;tt}^{t}=\frac{\dot{\gamma}(t)}{\gamma(t)}+\gamma(t),\quad
\Gamma_{\;tr}^{r}=\Gamma_{\;rt}^{r}=\Gamma_{\;t\theta}^{\theta}=\Gamma
_{\;\theta t}^{\theta}=\Gamma_{\;t\phi}^{\phi}=\Gamma_{\;\phi t}^{\phi}%
=\gamma(t), \label{con2}%
\end{equation}
where the dot denotes differentiation with respect to $t$.

Finally, the third connection, $\Gamma_{3}$, is characterized by the extra
nonzero components
\begin{equation}
\Gamma_{\;tt}^{t}=-\frac{\dot{\gamma}(t)}{\gamma(t)},\quad\Gamma_{\;rr}%
^{t}=\gamma(t),\quad\Gamma_{\;\theta\theta}^{t}=\gamma(t)r^{2},\quad
\Gamma_{\;\phi\phi}^{t}=\gamma(t)r^{2}\sin^{2}\theta. \label{con3}%
\end{equation}

In the case where $k\neq0$, there exist only one compatible connection with
nonzero coefficients%
\begin{equation}%
\begin{split}
&  \Gamma_{\;tt}^{t}=-\frac{k+\dot{\gamma}(t)}{\gamma(t)},\quad\Gamma
_{\;rr}^{t}=\frac{\gamma(t)}{1-kr^{2}}\quad\Gamma_{\;\theta\theta}^{t}%
=\gamma(t)r^{2},\quad\Gamma_{\;\phi\phi}^{t}=\gamma(t)r^{2}\sin^{2}(\theta)\\
&  \Gamma_{\;tr}^{r}=\Gamma_{\;rt}^{r}=\Gamma_{\;t\theta}^{\theta}%
=\Gamma_{\;\theta t}^{\theta}=\Gamma_{\;t\phi}^{\phi}=\Gamma_{\;\phi t}^{\phi
}=-\frac{k}{\gamma(t)},\quad\Gamma_{\;rr}^{r}=\frac{kr}{1-kr^{2}},\\
&  \Gamma_{\;\theta\theta}^{r}=-r\left(  1-kr^{2}\right)  , \quad
\Gamma_{\;\phi\phi}^{r}=-r\sin^{2}(\theta)\left(  1-kr^{2}\right)  ,
\quad\Gamma_{\;r\theta}^{\theta}=\Gamma_{\;\theta r}^{\theta}=\Gamma_{\;r\phi
}^{\phi}=\Gamma_{\;\phi r}^{\phi}=\frac{1}{r},\\
&  \Gamma_{\;\phi\phi}^{\theta}=-\sin\theta\cos\theta, \quad\Gamma
_{\;\theta\phi}^{\phi}=\Gamma_{\;\phi\theta}^{\phi}=\cot\theta.
\end{split}
\label{conk1}%
\end{equation}
We will refer to this last connection as $\Gamma^{(k)}$. We observe that when
$k=0$, the latter connection reduces to the third connection of the flat case,
that is $\Gamma^{(0)}=\Gamma_{3}$.

\section{Spatially flat case $k=0$}

\label{sec3}

Enforcing the homothetic restriction in the metric tensor (\ref{genlineel})
for the spatially flat case, i.e. $L_{\xi}g_{\mu\nu}= 2 g_{\mu\nu}$, where $L$
stands for the Lie derivative, we conclude that
\begin{equation}
\xi=t\partial_{t}+\left(  1-\lambda\right)  r\partial_{r}%
\end{equation}
is a homothetic vector field, and the corresponding line element reads%
\begin{equation}
ds^{2}=-dt^{2}+t^{2\lambda}\left(  dr^{2}+r^{2}\left(  d\theta^{2}+\sin
^{2}\theta d\phi^{2}\right)  \right)  . \label{sd1}%
\end{equation}
When $\lambda=0$, the flat space is recovered, thus we exclude this value from
our considerations.

We proceed to investigate the $f(Q)$ theories that admit such a
line element as a solution for the different connections we discussed in the
previous section. Note that, as previously stated, there exists a coordinate
system, where each of the connections becomes zero. However, this coordinate
system is different for each of the admitted connections and - in most
situations - leads to a transformed metric that loses its obvious homogeneity
and isotropy. Of the various connections, only the $\Gamma_{1}$ can be made
zero by going to Cartesian coordinates and by adopting a simple time
re-parametrization, thus, maintaining the line-element in its commonly
encountered form. Here we work in spherical coordinates, with the line element
given by \eqref{sd1} and consider the corresponding non-zero connections
compatible with this expression.

\subsection{The first connection $\Gamma_{1}$}

For the first connection the non-metricity scalar is derived
\begin{equation}
\label{Qvalg1}Q=-6\left(  \frac{\dot{a}}{a}\right)  ^{2}%
\end{equation}
thus, for the line element (\ref{sd1}) it follows
\begin{equation}
Q=-\frac{6{c}_{1}^{2}}{t^{2}}. \label{sd2}%
\end{equation}
Note that, with the definition \eqref{defQ} we used, the resulting $Q$ in the
case of $\Gamma_{1}$ results to be negative. This is purely a matter of
convention, in other works in the literature the opposite definition is
sometimes adopted.

The equations of motion for the connection (\ref{feq2}), are
identically satisfied. Thus, the connection $\Gamma_{1}$ plays no role in the
dynamics. This corresponds to the usual case studied in the literature; that
of the spatially flat FLRW geometry, with the metric taken in Cartesian
coordinates and the connection in the coincident gauge, $\Gamma^{\kappa
}_{\phantom{\kappa}\lambda\mu}=0$.

It is easily seen that the gravitational field equations are independent of
the function $\gamma(t)$ and they are
\begin{equation}
\frac{3\lambda^{2}f^{\prime}(Q)}{t^{2}}-\frac{1}{2}Qf^{\prime}(Q)+\frac{1}%
{2}f(Q)=\rho, \label{sd3}%
\end{equation}%
\begin{equation}
\lambda\left(  2-3\lambda\right)  t^{-2}f^{\prime}(Q)-2\lambda t^{-1}\dot
{Q}f^{\prime\prime}(Q)+\frac{1}{2}\left(  Q f^{\prime}(Q)-f(Q) \right)  =p,
\end{equation}
where we have considered the tensor $T_{\phantom{\mu}\nu}^{\mu}=\mathrm{diag}%
(-\rho(t),p(t),p(t),p(t))$ to describe the energy momentum tensor.

Assume now a perfect fluid with constant equation of state parameter, that is,
$p=w\rho$, thus the equation of motion for the matter fluid $T^{\mu}{}%
_{\nu;\mu}=0$, provides%
\begin{equation}
t\dot{\rho}+3\lambda\left(  1+w\right)  \rho=0,
\end{equation}
with exact solution $\rho\left(  t\right)  =\rho_{0}t^{-3\lambda\left(
1+w\right)  }$ where~$\rho_{0}$ is an integration constant. Under the above
substitutions, as well as eliminating explicit time dependence through
(\ref{sd2}), equation (\ref{sd3}) is transcribed to the following form
\begin{equation}
\rho_{0}\left(  6\lambda^{2}\right)  ^{\frac{-3\lambda(w+1)}{2}}
(-Q)^{\frac{3}{2}\lambda(w+1)}-Qf^{\prime}(Q)+\frac{1}{2}f(Q)=0. \label{sd4}%
\end{equation}

The latter equation is a first-order ordinary differential equation in the
dependent variable $f\left(  Q\right)  $, while $Q$ is the independent
variable. The exact solution of equation (\ref{sd4}) is expressed as follows
\begin{equation}
\label{fQtheor1}f(Q)=f_{0}\sqrt{-Q}+f_{1}(-Q)^{\frac{3}{2}\lambda(w+1)},
\end{equation}
where $f_{0}$ is an arbitrary constant, while $f_{1}$ is defined as%
\begin{equation}
f_{1}=\frac{2 \lambda^{-3 \lambda(1+w)} \rho_{0}}{6^{\frac{3}{2}\lambda
(w+1)}\left(  1 - 3\lambda\left(  1+w \right)  \right)  }.
\end{equation}
The minus signs are placed in Eq. \eqref{fQtheor1} in order to
have a real valued action for real constants of integration; remember that in
our conventions we obtain $Q<0$ in this section, see Eq. \eqref{Qvalg1}.

We observe that, self-similarity in the case of $\Gamma_{1}$, implies a
power-law type of $f(Q)$ theory where $Q$ is raised to the power $\frac{3}%
{2}\lambda(w+1)$. The constant $\lambda$ is the one appearing in the line
element \eqref{sd1}, while $w$ is the equation of state parameter. If we
consider the value which linearizes the relevant term in \eqref{fQtheor1},
that is $\frac{3}{2}\lambda(w+1)=1$, then we recover the solution of General
Relativity with $a(t)=t^{\lambda}=t^{\frac{2}{3(1+w)}}$.

The extra $\sqrt{-Q}$ term, that we see in \eqref{fQtheor1}, contributes just
as a surface term at the level of the action; see Eq. \eqref{Qvalg1}, the
$\sqrt{-Q}= \frac{d}{dt}\left(  \sqrt{6}\ln a\right)  $ is a total derivative
of a function involving the scale factor. In fact, if we consider the same
configuration in the absence of matter, $\rho=p=0$, the only solution which is
obtained is $f(Q)\sim\sqrt{-Q}$, which is trivial in the sense that, for
$\Gamma_{1}$, the $f(Q)\sim\sqrt{-Q}$ theory admits all metrics as solutions.

It is quite simple to repeat the above calculations for different equations of
state. If we consider a generic barotropic equation $p=p(\rho)$, then the
continuity condition $T^{\mu}{}_{\nu;\mu}=0$, results in
\begin{equation}
2 Q \frac{d\rho}{d Q} -3 \lambda(p(\rho(Q))+\rho(Q)) =0,
\end{equation}
where we have used \eqref{sd2} to make the change of variables $t\rightarrow
Q$. The latter can be integrated to give
\begin{equation}
\label{intarbcont}\int\!\! \frac{d\rho}{p(\rho)+\rho} = \frac{3 \lambda}{2}
\ln Q .
\end{equation}
However, whether it will be possible to invert \eqref{intarbcont}, in order to
obtain the $\rho(Q)$, which is to be used in \eqref{sd3} for the derivation of
a differential equation in the $f(Q)$, depends on the particular equation of
state, $p(\rho)$, under consideration. In any case, the \eqref{sd3} is going to
produce a first order ordinary differential equation, which is of the form
\begin{equation}
\frac{f(Q)}{2} - Q f^{\prime}(Q) = \rho(Q),
\end{equation}
solved by
\begin{equation}
\label{theorg1gen}f(Q) = \sqrt{-Q}\int\!\! \frac{\rho(Q)}{(-Q)^{\frac{3}{2}}}
dQ .
\end{equation}
For the case of the linear equation of state $p=w \rho$, relation
\eqref{intarbcont} leads to $\rho\sim(-Q)^{\frac{3}{2} \lambda(w+1)}$, and
subsequently through \eqref{theorg1gen} to the theory we obtained in \eqref{fQtheor1}.

We can now easily generalize solution \eqref{fQtheor1} by considering an
arbitrary number of perfect fluids with different equations of state,
$p_{i}=w_{i} \rho_{i}$. If we assume the sufficient (but not necessary)
condition that each fluid separately satisfies a continuity equation, so that
$\rho_{i} = C_{i} (-Q)^{\frac{3}{2} \lambda(w_{i}+1)}$, then, due to $\rho(Q)$
in \eqref{theorg1gen} being $\rho(Q) = \sum_{i=1}^{n} \rho_{i} $, we obtain a
theory of the form
\begin{equation}
f(Q) = f_{0} \sqrt{-Q} + 2 \sum_{i=1}^{n} \frac{C_{i} (-Q)^{\frac{3}{2}
\lambda(w_{i}+1)} }{1-3\lambda(1+w)}, \quad n\in\mathbb{Z}_{+},
\end{equation}
where the $C_{i}$ correspond to constants of integration. We have thus
obtained a series, comprised of terms involving the non-metricity scalar $Q$,
raised in powers associated with the equation of state parameter $w_{i}$ of
each fluid.

\subsection{The second connection $\Gamma_{2}$}

For the second connection the non-metricity scalar is obtained as
\begin{equation}
Q=-\frac{6\dot{a}^{2}}{a^{2}}+9\gamma\frac{\dot{a}}{a}+3\dot{\gamma}
\label{Qcon2}%
\end{equation}
where it is clear that function $\gamma\left(  t\right)  $ is involved. The
field equations in the case of vacuum are
\begin{subequations}
\label{feq12}%
\begin{align}
&  \frac{3\dot{a}^{2}f^{\prime}(Q)}{a^{2}}+\frac{1}{2}\left(  f(Q)-Qf^{\prime
}(Q)\right)  +\frac{3\gamma\dot{Q}f^{\prime\prime}(Q)}{2}=0,\label{feq12b}\\
&  -2\frac{d}{dt}\left(  \frac{f^{\prime}(Q)\dot{a}}{a}\right)  -\frac
{3\dot{a}^{2}}{a^{2}}f^{\prime}(Q)-\frac{1}{2}\left(  f(Q)-Qf^{\prime
}(Q)\right)  +\frac{3\gamma\dot{Q}f^{\prime\prime}(Q)}{2}=0,
\end{align}
while the one for the connection yields
\end{subequations}
\begin{equation}
\dot{Q}^{2}f^{\prime\prime\prime}(Q)+\left[  \ddot{Q}+\dot{Q}\left(
\frac{3\dot{a}}{a}\right)  \right]  f^{\prime\prime}(Q)=0. \label{feq22aa}%
\end{equation}
Note that, unlike what we saw in the previous section for connection
$\Gamma_{1}$, here the vacuum case is not trivial.

For the line element (\ref{sd1}), the scalar (\ref{Qcon2}) becomes
\begin{equation}
Q=-\frac{6\lambda^{2}}{t^{2}}+\frac{9\lambda\gamma}{t}+3\dot{\gamma}.
\label{defQ12}%
\end{equation}
The equations of motion for the connection (\ref{feq2}) reads%
\begin{equation}
3\lambda\dot{Q}f^{\prime\prime}(Q)+t\dot{Q}^{2}f^{\prime\prime\prime}\left(
Q\right)  +t\ddot{Q}(t)f^{\prime\prime}(Q) =0 \label{feq21}%
\end{equation}
and the gravitational field equations reduce to
\begin{equation}
\left(  \frac{6\lambda^{2}}{t^{2}}-Q\right)  f^{\prime}(Q)+3\gamma\dot
{Q}f^{\prime\prime}(Q)+f(Q) =0 \label{feq22a}%
\end{equation}%
\begin{equation}
\left(  -6\lambda^{2}+4\lambda+t^{2}Q\right)  f^{\prime}(Q)+t(3t\gamma\left(
t\right)  -4\lambda)\dot{Q} f^{\prime\prime}\left(  Q\right)  -t^{2}f(Q)=0 .
\label{feq22b}%
\end{equation}

%Solving the equation (\ref{defQ12}) in terms of $\dot{\gamma}$ it follows
%\begin{equation}
%\dot{\gamma}=\frac{\lambda(2\lambda-3t\gamma)}{t^{2}}+\frac{Q}{3}.\label{defQ'}%
%\end{equation}
By solving \eqref{feq22a} and \eqref{feq22b} algebraically with respect to
$\gamma$ and $f^{\prime\prime}(Q)$, we find:
\begin{subequations}
\label{feq33b}%
\begin{align}
\gamma &  =\frac{2\lambda\left(  \left(  6\lambda^{2}-t^{2}Q\right)
f^{\prime}\left(  Q\right)  +t^{2}f(Q)\right)  }{3t\left(  \left(
2\lambda(3\lambda-1)-t^{2}Q\right)  f^{\prime}(Q)+t^{2}f(Q)\right)  }\\
f^{\prime\prime}(Q)  &  =\frac{\left(  2\lambda(1-3\lambda)+t^{2}Q\right)
f^{^{\prime}}\left(  Q\right)  -t^{2}f(Q)}{2\lambda t\dot{Q}}.
\end{align}

%Consequently, with the use of the latter expression from the equation of
%motion (\ref{feq21}) we end with the following equation
%\begin{equation}
%f^{\prime\prime\prime}(Q(t))=\frac{\left(  \left(  2\lambda(3\lambda%
%-1)-t^{2}Q\right)  f^{\prime}\left(  Q\right)  +t^{2}f(Q)\right)  \left(
%3\lambda\dot{Q}+t\ddot{Q}\right)  }{2\lambda t^{2}Q^{\prime3}}\label{deff^3}%
%\end{equation}

Taking the time derivative of (\ref{feq33b}), solving it for $f^{\prime
\prime\prime}(Q)$ and substituting it together with \eqref{feq33b} in
\eqref{feq21} we arrive at the integrability condition
\end{subequations}
\begin{equation}
\left(  4\lambda+t^{2}Q\right)  f(Q)-Q\left(  t^{2}Q-6(\lambda-1)\lambda
\right)  f^{\prime}(Q)=0. \label{deff^2}%
\end{equation}
This relation provides a relation for the first derivative of $f(Q)$ as long
as $t^{2}Q-6(\lambda-1)\lambda\neq0$. But let us first examine the special
case, where $t^{2}Q-6(\lambda-1)\lambda=0$.

\subsubsection{Special case}

If we require $t^{2}Q-6(\lambda-1)\lambda=0$, then, as long as $f(Q)\neq0$,
Eq. \eqref{deff^2} implies $4\lambda+t^{2}Q=0$. These two conditions for $Q$
combined lead to the algebraic condition $(1-3\lambda)\lambda=0$, that is,
$\lambda=0$ or $\lambda=\frac{1}{3}$. The case $\lambda=0$ corresponds to the
flat space and leads to $Q=0$, so we concentrate our attention at the value
$\lambda=\frac{1}{3}$. For this $\lambda$ we have
\begin{equation}
\label{Qsp13}Q= -\frac{4}{3t^{2}} .
\end{equation}
By calculating $\dot{Q}$ and $\ddot{Q}$ and by inverting \eqref{Qsp13} to
substitute $t\rightarrow Q$ inside \eqref{feq21}, we arrive to a differential
equation for $f(Q)$:
\begin{equation}
Qf^{\prime\prime\prime}(Q)+f^{\prime\prime}(Q)=0.
\end{equation}
Its solution is $f\left(  Q\right)  =f_{1}\, Q+f_{2} \, Q\ln(- Q) +f_{3} $,
with the $f_{1}$, $f_{2}$ and $f_{3}$ being constants of integration. In
addition, since we know $Q$ from \eqref{Qsp13}, we can use it in
\eqref{defQ12} and integrate the latter to obtain the function which enters
the connection $\Gamma_{2}$. Thus, we derive
\begin{equation}
\label{gammasp13}\gamma= \frac{C}{t} -\frac{2}{9 t}\ln t,
\end{equation}
where $C$ is a constant of integration. The above relations \eqref{Qsp13},
\eqref{gammasp13}, with the derived expression for $f(Q)$ satisfy the field
equations under the following conditions for the constants of integration
\begin{equation}
f_{3}=0 \quad\text{and} \quad C= \frac{f_{1} + f_{2}\left(  3+\ln\left(
\frac{4}{3}\right)  \right)  }{9 f_{2}} .
\end{equation}
Thus, the theory we obtain in this case is
\begin{equation}
f\left(  Q\right)  =f_{1}\, Q+f_{2} \, Q\ln\left(  - Q\right)  ,
\end{equation}
which introduces a logarithmic modification to GR, $f(Q)\sim Q$. Notice that
the minus sign enters the logarithm due to the $Q$ of \eqref{Qsp13} being
negative definite.

\subsubsection{General case}

Leaving aside the above special case, if we now consider $t^{2}Q-6(\lambda
-1)\lambda\neq0$, then equation \eqref{deff^2} gives
\begin{equation}
f^{\prime}(Q)=\frac{\left(  4\lambda+t^{2}Q\right)  f(Q)}{Q\left(
t^{2}Q-6(\lambda-1)\lambda\right)  }. \label{deff^1}%
\end{equation}
We can take the time derivative of the above expression and solve with respect
to $f^{\prime\prime}(Q)$, then we substitute this relation together with
\eqref{deff^1} inside \eqref{feq33b} and its time derivative. By substituting
the latter two in \eqref{defQ12}, we arrive at a differential equation for
$Q$
\begin{equation}
\left(  1-3 \lambda\right)  Q + t \dot{Q} =0,
\end{equation}
which yields
\begin{equation}
\label{Qgeng2}Q = q_{0}t^{3\left(  \lambda-1\right)  },
\end{equation}
where $q_{0}$ is a constant. With the use of the above expression in
\eqref{defQ12} we can integrate the latter to obtain the corresponding
connection which results in
\begin{equation}
\gamma= \frac{q_{0} t^{3 \lambda}+12 \lambda^{2} t}{6 (3 \lambda-1) t^{2}}+ C
t^{-3 \lambda},
\end{equation}
with $C$ denoting once more a constant of integration. By calculating the
derivatives of \eqref{Qgeng2} and by also inverting \eqref{Qgeng2} to obtain a
mapping $t\rightarrow Q$ we obtain from the connection equation \eqref{feq21}
the following condition on $f(Q)$:
\begin{equation}
3 (\lambda-1) Q f^{\prime\prime\prime}(Q)+2 (3 \lambda-2) f^{\prime\prime}(Q)
=0,
\end{equation}
which is solved by
\begin{equation}
f(Q) = f_{1} Q + f_{2} Q^{\frac{2}{3(1-\lambda)}} + f_{3},
\end{equation}
with $f_{1}$, $f_{2}$ and $f_{3}$ denoting the integration constants.
{However, we need to substitute the above acquired expressions in
the rest of the field equations \eqref{feq22a} and \eqref{feq22b}. When we do
so we observe that the following conditions must be set
\begin{equation}
C=f_{3}=0 \quad\text{and} \quad f_{2}= -6 f_{1} (\lambda-1) \lambda
q_{0}^{\frac{2}{3 (\lambda-1)}},
\end{equation}
which results in a theory characterized by
\begin{equation}
f(Q) = f_{1} \left(  Q -6 (\lambda-1) \lambda q_{0}^{\frac{2}{3 (\lambda-1)}}
Q^{\frac{2}{3(1-\lambda)}} \right)  .
\end{equation}
Thus, in the generic case $\lambda\neq\frac{1}{3}$ we get a theory which
involves a power-law modification to General Relativity. The case $\lambda=1$
corresponds to a GR solution and yields $Q=$const.}

%We proceed now by considering the time derivative of \eqref{feq33b} and
%equating with expression \eqref{defQ'} (using of course the available rules
%for the derivatives), we arrive at the integrability condition \eqref{eqalG}
%\begin{equation}
%-\frac{\left(  12\lambda^{2}+(2-3\lambda)t^{2}Q\right)  \left(  \dot{Q}-1\right)
%}{6(3\lambda-1)t^{2}\dot{Q}}=0\label{eqalG}%
%\end{equation}

%We repeat the same procedure for the equations \eqref{deff^1} and
%(\ref{feq22a,feq22b}) we obtain the second integrability condition
%\eqref{eqalf(Q)}
%\begin{equation}
%\frac{8\lambda^{2}(3\lambda-1)f(Q)\left(  3(\lambda-1)Q-t\dot{Q}\right)  }%
%{tQ^{2}\left(  t^{2}Q-6(\lambda-1)\lambda\right)  ^{2}\dot{Q}}=0\label{eqalf(Q)}%
%\end{equation}

%The above 2 conditions result in various cases of study. We present only the
%cases for the free parameters where accepted solutions exist.

%\textbf{ }For $3(\lambda-1)Q(t)-tQ^{\prime}(t)=0$, the solution being
%$Q(t)=q_{0}t^{3\left(  \lambda-1\right)  }$. If we invert this relation and
%eliminate the explicit to the time-dependence of \eqref{deff^1} we arrive at a
%differential equation of $f(Q(t))$ which is solved by the closed-form
%expression
%\begin{equation}
%f(Q(t))=f_{1}Q(t)^{\frac{2}{3\left(  1-\lambda\right)  }}+f_{2}%
%Q(t).\label{ai.03}%
%\end{equation}

%On the other hand, for $\lambda=0$, which describes the four-dimensional flat
%spacetime, from the field equations we derive $f\left(  Q\right)  =f_{1}Q$.

\subsection{The third connection $\Gamma_{3}$}

The non-metricity scalar is calculated
\begin{equation}
Q=-\frac{6\dot{a}^{2}}{a^{2}}+\frac{3\gamma}{a^{2}}\frac{\dot{a}}{a}%
+\frac{3\dot{\gamma}}{a^{2}}, \label{Qcon3}%
\end{equation}
while the equations of motion for the metric in the case of vacuum are%
\begin{equation}
\frac{3\dot{a}^{2}f^{\prime}(Q)}{a^{2}}+\frac{1}{2}\left(  f(Q)-Qf^{\prime
}(Q)\right)  -\frac{3\gamma\dot{Q}f^{\prime\prime}(Q)}{2a^{2}}=0,
\end{equation}%
\[
-2\frac{d}{dt}\left(  \frac{f^{\prime}(Q)\dot{a}}{a}\right)  -\frac{3\dot
{a}^{2}}{a^{2}}f^{\prime}(Q)-\frac{1}{2}\left(  f(Q)-Qf^{\prime}(Q)\right)
+\frac{\gamma\dot{Q}f^{\prime\prime}(Q)}{2a^{2}}=0
\]
and for the connection we have
\begin{equation}
\dot{Q}^{2}f^{\prime\prime\prime}(Q)+\left[  \ddot{Q}+\dot{Q}\left(
\frac{\dot{a}}{a}+\frac{2\dot{\gamma}}{\gamma}\right)  \right]  f^{\prime
\prime}(Q)=0. \label{feq23}%
\end{equation}

For the self-similar line element (\ref{sd1}) we calculate
\begin{equation}
Q(t)=t^{-2(\lambda+1)}\left(  -6\lambda^{2}t^{2\lambda}+3\lambda
t\gamma+3t^{2}\dot{\gamma}\right)  . \label{defQ2}%
\end{equation}
or equivalently%
\begin{equation}
\dot{\gamma}=\frac{1}{3}t^{2\lambda}Q+\frac{\lambda\left(  2\lambda
t^{2\lambda}-t\gamma\right)  }{t^{2}}. \label{def1}%
\end{equation}

The field equations are written in the equivalent form%

\begin{equation}
\left(  \frac{6\lambda^{2}}{t^{2}}-Q\right)  f^{\prime}\left(  Q\right)
-3t^{-2\lambda}\gamma Q^{\prime}f^{\prime\prime}(Q)+f(Q) =0, \label{feq55a}%
\end{equation}
\begin{equation}
2\left(  \frac{\gamma}{4}-\lambda t^{2\lambda-1}\right)  \dot{Q}%
f^{\prime\prime}(Q)-\frac{1}{2}t^{2\lambda}\left(  f(Q)-Qf^{\prime}(Q)\right)
-\lambda(3\lambda-2)t^{2\lambda-2}f^{\prime}(Q)=0, \label{feq55b}%
\end{equation}
and%

\begin{equation}
\gamma\left(  \lambda\dot{Q}f^{\prime\prime}\left(  Q\right)  +t\ddot
{Q}f^{\prime\prime}\left(  Q\right)  +t\dot{Q}^{2}f^{\prime\prime\prime
}\left(  Q\right)  \right)  +2t\dot{Q}\dot{\gamma}f^{\prime\prime}\left(
Q\right)  =0. \label{ee1}%
\end{equation}

We notice that if we divide the equation of motion for the connection
(\ref{ee1}) with $t\gamma\dot{Q}f^{\prime\prime}(Q)$ we can infer an integral
of motion
\begin{equation}
t^{\lambda}\gamma^{2}\dot{Q}f^{\prime\prime}(Q)=m,
\end{equation}
that is
\begin{equation}
f^{\prime\prime}(Q)=\frac{mt^{-\lambda}}{\gamma^{2}\dot{Q}} ,
\label{detf''(Q)}%
\end{equation}
where we assume $m\neq0$ to avoid the linear $f(Q)\sim Q$ case. We replace
\eqref{detf''(Q)} into \eqref{feq55a}, \eqref{feq55b} and we find the
following relations%
\begin{equation}
f^{\prime}(Q)=\frac{t^{2}\left(  \gamma f(Q)-3mt^{-3\lambda}\right)  }%
{\gamma\left(  t^{2}Q-6\lambda^{2}\right)  }, \label{feq44b}%
\end{equation}%
\begin{equation}
f(Q)=\frac{mt^{-3\lambda-1}\left(  t^{2}Q\left(  2\lambda t^{2\lambda}%
+t\gamma\right)  -6\lambda\left(  2\lambda^{2}t^{2\lambda}+(\lambda
-1)t\gamma\right)  \right)  }{2\lambda\gamma^{2}}. \label{feq44a}%
\end{equation}

If we take the first derivative with respect to the time \eqref{feq44b} and
divide it by $\dot{Q}$ the result must be the same as in \eqref{feq44a}. We
thus end up with the equation
\begin{equation}
\left(  6\lambda^{2}-t^{2}Q\right)  \left(  6\left(  4\lambda^{3}t^{4\lambda
}+(\lambda-1)t^{2}\gamma^{2}\right)  +t^{2\left(  \lambda+1\right)  }Q\left(
4\lambda t^{2\lambda}+t\gamma\right)  \right)  =0. \label{def2}%
\end{equation}
The above relation allows for an algebraic derivation of $\gamma(t)$ with
respect to $t$ and $Q$. The subsequent derivation of $Q(t)$ by integrating
\eqref{defQ2} is straightforward. However, the resulting $Q(t)$ relation
cannot in general be inverted to obtain the mapping $t\rightarrow Q$, which
would allow us to write a differential equation for $f(Q)$ from the field
equations and thus obtain the general family of theories allowing for
self-similar solutions.

In order to proceed and obtain an $f(Q)$ theory, given in terms of elementary
functions, at least as a partial solution, we follow a different procedure.
Specifically, we make use of the existence of the homothetic vector field and
use it to set an additional restriction on the connection; In a sense, demand
self-similarity for the connection as well, not just for the metric. Due to
the fact that the connection is not a tensor, we choose to set a
\textquotedblleft homothetic\textquotedblright\ condition over the
non-metricity tensor and demand $L_{\text{$\xi$}}Q_{\lambda\mu\nu}=2\sigma
Q_{\lambda\mu\nu}$, where $\sigma$ is a constant. Interestingly enough, this
equation is solved if the homothetic factor is the same as that of the metric,
i.e. $\sigma=1$, while $\gamma$ satisfies the differential equation
\begin{equation}
\left(  2\lambda-1\right)  \gamma-t\dot{\gamma}=0, \label{homQddd}%
\end{equation}
that is, $\gamma=\gamma_{0}t^{2\lambda-1}$. However, for the compatable
connections which we consider in this work condition $L_{\text{$\xi$}%
}Q_{\lambda\mu\nu}=2Q_{\lambda\mu\nu}$ is equivalent with the condition
$L_{\xi}\Gamma_{~\mu\nu}^{\lambda}=0$. Thus we demand the autoparallels, that
is, the equations of motion for a test particle, to be invariant under the
action of the homothetic vector field as in the case of the Levi-Civita connection.

With this $\gamma(t)$, the non-metricity scalar reads%
\begin{equation}
Q(t)=-\frac{3\left(  2\lambda^{2}-3\lambda\text{$\gamma$}_{0}+\gamma
_{0}\right)  }{t^{2}}. \label{c01}%
\end{equation}
The field equations are now written as
\begin{equation}
\left(  6\lambda^{2}-t^{2}Q\right)  f^{\prime}(Q)-3\gamma_{0}t\dot{Q}%
f^{\prime\prime}+t^{2}f(Q)=0, \label{CH}%
\end{equation}%
\begin{equation}
\left(  -6\lambda^{2}+4\lambda+t^{2}Q\right)  f^{\prime}(Q)+t(\gamma
_{0}-\lambda)\dot{Q}(t)f^{\prime\prime}-t^{2}f(Q)=0,
\end{equation}
and%
\begin{equation}
\left(  (5\lambda-2)\dot{Q}f^{\prime\prime}(Q)+tQ^{\prime2}f^{\prime
\prime\prime}\left(  Q\right)  +t\ddot{Q}f^{\prime\prime}(Q)\right)  =0.
\label{MI}%
\end{equation}
By using (\ref{c01}) to convert equation (\ref{MI}) into a differential
equation for $f(Q)$, we obtain
\[
5\left(  \lambda-1\right)  f^{\prime\prime}\left(  Q\right)  -2Qf^{\prime
\prime\prime}\left(  Q\right)  =0,
\]
from where it follows%
\begin{equation} \label{fqsomesol}
f\left(  Q\right)  = f_{1} Q^{\frac{5\lambda-1}{2}}+f_{2}Q+f_{3}%
\end{equation}
with $\lambda\neq\frac{1}{5}$ and $\lambda\neq\frac{3}{5}$.

If we replace \eqref{fqsomesol} in the field equations, we find that the flat spacetime is
recovered, that is, $\lambda=0$, when $f_{3}=f_{1}=0$, which corresponds to GR
since $f(Q)\sim Q$. On the other hand, for $f_{2}=f_{3}=0$, i.e. a power-law $f(Q)$, there exists an
analytic solution with the constraint $\gamma_{0}=\frac{2\lambda\left(
2-5\lambda\right)  }{5\lambda-3}$. Finally, for $\lambda=\frac{1}{5}$ or
$\lambda=\frac{3}{5}$ there is not any valid solution for a function $\gamma$
obtained under the condition \eqref{homQddd}.

\section{Non-zero spatial curvature $k\neq0$}

\label{sec4}

Imposing the homothetic constraint, $L_{\xi}g_{\mu\nu}=2g_{\mu\nu}$, for the
case of non-zero spatial curvature we end with the line element%
\begin{equation}
ds^{2}=-dt^{2}+\left(  a_{0}t\right)  ^{2}\left[  \frac{dr^{2}}{1-kr^{2}%
}+r^{2}\left(  d\theta^{2}+\sin^{2}\theta d\phi^{2}\right)  \right]  ,
\end{equation}
where the homothetic vector field is $\xi=t\partial_{t}.$

For a non-zero spatial curvature $k$, the non-metricity scalar becomes
\begin{equation}
Q=-\frac{6\dot{a}^{2}}{a^{2}}+\frac{3\gamma}{a^{2}}\left(  \frac{\dot{a}}%
{a}\right)  +\frac{3\dot{\gamma}}{a^{2}}+k\left[  \frac{6}{a^{2}}+\frac
{3}{\gamma}\left(  \frac{\dot{\gamma}}{\gamma}-\frac{3\dot{a}}{a}\right)
\right]  . \label{Qconk}%
\end{equation}
The gravitational field equations in vacuum are
\begin{subequations}
\label{feq1k}%
\begin{align}
&  \frac{3\dot{a}^{2}f^{\prime}(Q)}{a^{2}}+\frac{1}{2}\left(  f(Q)-Qf^{\prime
}(Q)\right)  -\frac{3\gamma\dot{Q}f^{\prime\prime}(Q)}{2a^{2}}+3k\left(
\frac{f^{\prime}(Q)}{a^{2}}-\frac{\dot{Q}f^{\prime\prime}(Q)}{2\gamma}\right)
=0,\label{feq1kb}\\
&  -2\frac{d}{dt}\left(  \frac{f^{\prime}(Q)\dot{a}}{a}\right)  -\frac
{3\dot{a}^{2}}{a^{2}}f^{\prime}(Q)-\frac{1}{2}\left(  f(Q)-Qf^{\prime
}(Q)\right)  +\frac{\gamma\dot{Q}f^{\prime\prime}(Q)}{2a^{2}}\nonumber\\
&  -k\left(  \frac{f^{\prime}(Q)}{a^{2}}+\frac{3\dot{Q}f^{\prime\prime}%
(Q)}{2\gamma}\right)  =0,
\end{align}
and the equation of motion for the connection assumes the form
\end{subequations}
\begin{equation}
\dot{Q}^{2}f^{\prime\prime\prime}(Q)\left(  1+\frac{ka^{2}}{\gamma^{2}%
}\right)  +\left[  \ddot{Q}\left(  1+\frac{ka^{2}}{\gamma^{2}}\right)
+\dot{Q}\left(  \left(  1+\frac{3ka^{2}}{N^{2}\gamma^{2}}\right)  \frac
{\dot{a}}{a}+\frac{2\dot{\gamma}}{\gamma}\right)  \right]  f^{\prime\prime
}(Q)=0. \label{feq2k}%
\end{equation}

Once more, the problem is too complex to proceed without setting some
restriction on the function $\gamma(t)$. We try the same ansatz as in the
$k=0$ case of $\Gamma_{3}$ in the previous section, $L_{\text{$\xi$}%
}Q_{\lambda\mu\nu}=2Q_{\lambda\mu\nu}$, which can be seen to be equivalent to
$L_{\text{$\xi$}}\Gamma^{\lambda}{}_{\mu\nu}=0$. One of the non zero
independent terms of $L_{\text{$\xi$}}\Gamma^{\lambda}{}_{\mu\nu}=0$ is
$ \gamma(t)-t\dot{\gamma}(t)  =0$. The solution is
$\gamma(t)=\gamma_{0}t$, with $\gamma_{0}=$const., which satisfies the entire
homothetic equation. Thus, by replacing it in (\ref{Qconk}), we find%
\begin{equation}
Q(t)=-\frac{6\left(  a_{0}^{2}-\text{$\gamma$}_{0}\right)  (\gamma_{0}%
+k)}{a_{0}^{2}\gamma_{0}t^{2}}. \label{A2}%
\end{equation}
The field equations are%
\begin{equation}
\gamma_{0}\left(  6\left(  a_{0}^{2}+k\right)  -a_{0}^{2}t^{2}Q\right)
f^{\prime}(Q)-3t\left(  a_{0}^{2}k+\gamma_{0}^{2}\right)  \dot{Q}%
f^{\prime\prime}(Q)+ a_{0}^{2} \gamma_{0}t^{2}f(Q)=0, \label{feq66a}%
\end{equation}%
\begin{equation}
t\left(  a_{0}^{2}(4\gamma_{0}+3k)-\gamma_{0}^{2}\right)  \dot{Q}%
)f^{\prime\prime}(Q)+\gamma_{0}\left(  2\left(  a_{0}^{2}+k\right)  -a_{0}%
^{2}t^{2}Q\right)  f^{\prime}(Q)+a_{0}^{2}\gamma_{0}t^{2}f(Q)=0,
\label{feq66b}%
\end{equation}
and
\begin{equation}
3\left(  a_{0}^{2}k+\gamma_{0}^{2}\right)  \left(  tf^{\prime\prime\prime
}(Q)\dot{Q}^{2}+t\ddot{Q}f^{\prime\prime}(Q)+3\dot{Q}f^{\prime\prime
}(Q)\right)  =0. \label{NI}%
\end{equation}

Hence, with the use of (\ref{A2}) in (\ref{NI}) and given $Q\neq 0$, it follows%
\begin{equation}
\left(  a_{0}^{2}k+\gamma_{0}^{2}\right)  f^{\prime\prime\prime}\left(
Q\right)  =0.
\end{equation}

For $\left(  a_{0}^{2}k+\gamma_{0}^{2}\right)  \neq0$, we find $f\left(
Q\right)  =f_{1}+f_{2}Q+f_{3}Q^{2}$, and by replacing it in the field equations,
for arbitrary $k$, we end up with the constraints $f_{1}=f_{2}=0$ and
$\gamma_{0}=-3a_{0}^{2}$. There is also the special case  $\gamma_{0}=a_{0}^{2}$, which however leads to $Q=0$. For a constant $Q$ the equation of the connection is identically zero and sets no restriction in the functional form of $f(Q)$. Through the rest of the equations of motion, we see that, the $\gamma_{0}=a_{0}^{2}$, $Q=0$ case allows for two possibilities: either $k=-a_0^2=-1$ (Milne universe) or $a_0$ arbitrary and $k = \pm 1$; the first requires $f(0)=0$, while the second leads to the constraint $f(0)=f'(0)=0$.

On the other hand, for $k=-1$ and $\gamma_{0}=\pm a_{0}$, from (\ref{A2}) and
(\ref{feq66a}) it follows that $Q=-\frac{6\left(  a_{0} \mp1\right)  ^{2}%
}{a_{0}^{2} t^{2}}$. From equations \eqref{feq66a} and \eqref{feq66b} we
further obtain
\begin{equation}
f(Q) = q_{0} Q^{\frac{a_{0} \mp1}{2a_{0}}},
\end{equation}
where $q_{0}$ is a constant and it is required that $a_{0}\neq\pm1$. The
$a_{0} = \mp1$ case leads to $f(Q)\sim Q$, which leads to the Milne universe
solution with $k=-1$.

We briefly summarize the results we obtained in this and in the previous
sections, for the various types of spatial geometry and the corresponding
connections, in table \ref{table1}.

\begin{table}[ptb]%
\begin{tabular}
[c]{|c|c|c|c|c|c|}\hline
\makecell{spatial \\ curvature} & \multicolumn{3}{|c|}{$k=0$} & $k=+1$ &
$k=-1$\\[0.5ex]\hline
connection & $\Gamma_{1}$ & $\Gamma_{2}$ & $\Gamma_{3}$ & $\Gamma^{(+1)}$ &
$\Gamma^{(-1)}$\\[0.5ex]\hline
$f(Q)$ & $\sim|Q|^{\frac{3}{2}\lambda(w+1)}$ &
\makecell{$f_1 Q + f_2 Q \ln|Q|, \quad \lambda=\frac{1}{3}$\\ or \\ $ f_1 Q + f_2 Q^{\frac{2}{3(1-\lambda)}},$} &
$\sim Q^{\frac{5\lambda-1}{2}}$ & $\sim Q^{2}$ &
\makecell{$\sim Q^{2}, \; \forall \; a_0 \in \mathbb{R}^*$ \\ or \\ $\sim Q^{\frac{a_0\mp1}{2a_0}}$}\\\hline
matter content & $p=w \rho$ & - & - & - & -\\\hline
\end{tabular}
\caption{The functional form of the $f(Q)$ theory resulting in self-similar
gravitation field for the various types of spatial geometry and connections.
We include in the table only the cases consisting of modifications of GR.
Notice, that for the last three cases ($\Gamma_{3}$ and the two $k\neq0$), the
$f(Q)$ function is derived under the extra condition of self-similarity over
the non-metricity tensor.}%
\label{table1}%
\end{table}

\section{Conclusions}

\label{sec5}

In this study we investigate the existence of self-similar solutions in
$f\left(  Q\right)  $-symmetric teleparallel gravitational theory for four
distinct families of connections related with a homogeneous and isotropic
background geometry described by the FLRW line element. Three of the distinct
connections describe spatially flat FLRW spacetimes, while the fourth
connection corresponds to a non-zero spatial FLRW geometry. For each family of
connection we present the gravitational field equations and the equation of
motion for the connection. We assume that the background geometry admits an
homothetic vector field such that to be a self-similar spacetime. Hence, for
the exact functional form of the dynamical variables of the metric we solve
the field equations in order to constraint the connection and the $f\left(
Q\right)  $-function.

For the first connection we observe that it is necessary to introduce an
external matter source such that the self-similar solution to exist. The
resulting $f\left(  Q\right)  $-function is of power-law and it can be
extended into a series by adding different fluids in the matter sector. For
the second and third families of connections it is not necessary to introduce
an external matter source, and for power-law $f\left(  Q\right)  $ and
logarithmic modifications to GR self-similar solutions exist. Finally, for the
fourth connection which corresponds to a universe with a non-zero spatial
curvature, self-similar solutions exist for $f\left(  Q\right)  =Q^{2}$, when
the curvature is positive, or for a general power-law $f\left(  Q\right)  $
theory when the space is hyperbolic. For the latter case we also
demonstrated the consistency of the equations by obtaining the Milne universe,
when $f(Q)\sim Q$.

As we know, the general relativistic limit in the case of symmetric teleparallel gravity lies in the linear theory $f(Q)\sim Q$. If we purely concentrate on the form of the action, most of the reconstructed $f(Q)$ theories, which are seen in table \ref{table1}, can approach this limit by constraining appropriately the remaining free parameters. It is particularly interesting the case of connection $\Gamma_2$, where this limit can be directly achieved by simply requiring $f_2\rightarrow 0$. In the rest of the cases, a constant which enters the space-time metric (either $\lambda$ or $a_0$) needs to be constrained, but the limit is still achievable.

The use of the different connections in this work reveals additional applicable dynamics. We see how the same gravitational field $g_{\mu\nu}$ is produced by different $f(Q)$ theories depending on the type of connection you introduce. However, the physical interpretation of the different connections is not obvious, especially if you rely on observed quantities whose value depends purely on the metric, e.g. the Hubble function. Possible observable effects owed to the non-metricity have been explored in \cite{Latorre1} by considering matter which couples to the connection (e.g. fermions). However, even in this setting, it has been shown that the result depends on the level where you introduce the non-Riemannian modification \cite{Latorre2}. If you start by modifying the Dirac equation, then the non-metricity introduces an extra coupling among the fermions. On the other hand, if you insert the modification at the level of the action, the effect of the non-metricity disappears from the resulting field equation. In the first case, for a fixed metric $g_{\mu\nu}$, the differences owed to the distinct admissible $\Gamma^{\kappa}_{\;\lambda\mu}$, characterized by the $\gamma(t)$, could, in principle, be quantified.

The importance of self-similar spacetimes is well established in General Relativity \cite{Eardley}. They form simple solutions which however play a significant role as asymptotic limits of more complicated gravitational configurations, for example consider the role of the Kasner solution in the application of the Belinski–Khalatnikov–Lifshitz (BKL) conjecture \cite{BKL}. At this point it is difficult to evaluate if self-similarity will prove as important and as general in $f(Q)$ theory. But, given the fact that the field equations of the theory can be formulated as that of an effective fluid in General Relativity, it is expected that, at least under specific configurations, self-similarity will continue playing a role in limiting cases. From the analysis performed in previous works \cite{ww3,nfq} in FLRW spacetimes and for the case of the first connection we study here, it appears that self-similar cosmological solutions do indeed describe the asymptotic limits of more general solutions. A similar conclusion with that of
General Relativity. For the rest of the connections and for the function form
of the non-metricity scalar we do the hypothesis that a similar conclusion is
valid. However, that should be investigated further. Lastly, partial results from ongoing work of ours show that the general connections, compatible to the symmetries of a chosen background geometry, can be retrieved by the mere knowledge of the corresponding flat, self-similar connections.

In a future work we plan to investigate if the latter conclusion is valid in
the case of anisotropic geometries by investigating self-similar solutions in
Bianchi spacetimes.

\begin{acknowledgments}
N. D. acknowledges the support of the Fundamental Research Funds for the
Central Universities, Sichuan University Full-time Postdoctoral Research and
Development Fund No. 2021SCU12117. A.P. was supported in part by the National
Research Foundation of South Africa (Grant Numbers 131604).
\end{acknowledgments}

\end{document}